\documentclass[float,reprint,amsmath,amssymb,aps,superscriptaddress,nofootinbib,twocolumn,prd]{revtex4-2}
\usepackage{graphicx}
\usepackage{dcolumn}
\usepackage{bm}
\usepackage[utf8]{inputenc}
\usepackage{amsmath, amsthm, amsfonts, amssymb}
\usepackage[svgnames]{xcolor}
\definecolor{DarkRed}{RGB}{179, 27, 27} 
\colorlet{color1}{NavyBlue}
\usepackage[colorlinks=true,allcolors=DarkRed]{hyperref}
\usepackage{bm}
\usepackage{physics}
\usepackage{dsfont}
\usepackage{graphicx}
\usepackage{cleveref}
\usepackage{xfrac}
\usepackage{mathrsfs}
\usepackage{comment}

\definecolor{darkred}{rgb}{0.5,0,0}
\definecolor{darkgreen}{rgb}{0,0.4,0}
\definecolor{darkblue}{rgb}{0,0,0.5}

\def\apj{{ApJ}}

\def\apjl{{ApJL}}

\def\aap{{A\&A}}

\def\mnras{{MNRAS}}

\def\prd{{Physical Review D}}
\def\prl{{Phys. Rev. Lett.}}
\def\pre{{Physical Review E}}

\def\04a{{2004 a}}
\def\04b{{2004 b}}

\newcommand{\bt}{\boldsymbol{\tau}_{\rm el}}
\newcommand{\bb}{\boldsymbol{B}}
\newcommand{\bv}{\boldsymbol{v}}
\newcommand{\bnabla}{\boldsymbol{\nabla}}
\newcommand{\bxi}{\boldsymbol{\xi}}
\newcommand{\oS}{\omega_{\rm S}}
\newcommand{\oA}{\omega_{\rm A}}

\newcommand{\Rstar}{R}
\newcommand{\Ro}{\textrm{Ro}}

\begin{document}

\title{Magnetorotational instabilities in solids: Application to neutron-star crusts}

\author{Arthur G. Suvorov}
\affiliation{Theoretical Astrophysics, Institute for Astronomy and Astrophysics, University of T\"{u}bingen, 72076 T\"{u}bingen, Germany}
\author{Thomas Celora}
\affiliation{Institut de Ciències de I'Espai (ICE-CSIC), Campus UAB, 08193 Cerdanyola del Vallès, Barcelona, Spain}
\affiliation{Institut d’Estudis Espacials de Catalunya (IEEC), 08860 Castelldefels, Barcelona, Spain}
\author{Kostas D. Kokkotas}
\affiliation{Theoretical Astrophysics, Institute for Astronomy and Astrophysics, University of T\"{u}bingen, 72076 T\"{u}bingen, Germany}

\begin{abstract}
\noindent The magnetorotational instability can generate strong, turbulent substructure within magnetized shear flows.
The efficacy of the mechanism as a function of microphysical aspects of the fluid, such as stratification and diffusivity, has been explored extensively.
One aspect that has not been studied thus far, however, is whether the instability can also operate in solids.
Motivated by the possibility that solid regions within planets or degenerate stars may rotate differentially with respect to liquid or gaseous layers during some phase of their life, we examine the extent to which elasticity suppresses the instability. 
A simplified, plane-parallel analysis reveals that only in cases where the flow is strongly sheared, such that the magnetic tension that would result from the instability in a liquid exceeds the shear modulus of the elastic cavity, can magnetic growth occur.
In the context of dynamical tides in binary neutron-star mergers, this implies that the magnetic field can be amplified in the crust prior to coalescence only if the star boasts a spin frequency of $\gtrsim 300$~Hz.
If viscous heating weakens the crystalline structure prior to resonance, the required spin frequency is reduced.
\end{abstract}

\maketitle

\section{Introduction} \label{sec:intro}

Magnetised fluids hosting shear flows are subject to a wealth of magnetohydrodynamic (MHD) instabilities.
One of the more well-studied examples is the magneto-rotational instability (MRI) where a flow, if sheared in directions orthogonal to gradients of magnetic tension, can generate small-scale magnetic substructures \cite{veli59,c60}.
This mechanism allows for turbulence to be sustained over secular timescales, and is therefore thought to be crucial in mediating angular-momentum transport in astrophysical systems like accretion discs \cite{bal91,bal98}, stellar interiors \cite{spruit99,grif22}, jets \cite{og13,haw15}, and even molecular clouds \cite{kim03,pio07}.
Nature is complicated, however, and many restorative and diffusive ingredients can affect MRI activation criteria and saturation. 
The impact of buoyancy \cite{guil15,held24}, magnetic topology \cite{bai13,cel25}, general-relativistic effects \cite{duez06,eti06}, and chemical/Ohmic/thermal diffusivity \cite{ag78,corpi10} has been studied extensively.
One effect that has not been explored to date, however, is elasticity. 

In a solid, restorative forces set by chemical or ionic bonds resist stretching and compression.
This makes it difficult to sustain shear flows, which is likely why MRI activation in a solid has not received much attention.
However, transient shear flows can manifest in the astrophysical context through pulsations: degenerate stars \cite{xu03,horo10} or planets \cite{boden86,fuller14} with solid layers may have their non-radial oscillation modes excited through tides, accretion, or other mechanisms.
If the eigenfunctions host no nodes in the solid layer, bulk shear flows may operate there as the local angular velocity, $\Omega$, is dragged every half period.
Depending on the mode phase we thus expect segments of time where the radial gradient is negative, $\partial  \Omega / \partial r <0$, which is precisely the classical criterion for the onset of the MRI \cite{bal98}.
The diffusive, inward propagation of torques applied to the surface could also allow for continuous lags to build up in neutron stars \cite{mel12,glamp15,anz20}.
It is tempting therefore to conjecture that the MRI could still function in some cases.
In this paper, using a local plane-wave analysis as in the classical work of \citet{bal91}, we explore how the elasticity of a solid cavity interacts with any would-be instability. 

In the problem of binary neutron-star mergers, dynamical tides can drive modes to large amplitudes through orbital resonances \cite{shib94,lai94,kokk95,ho99}.
This led \citet{akrk24} (henceforth SKRK24) to study mode-induced magnetic amplification via the MRI, where it was found that magnetic fields of order $B \gtrsim 10^{13}$~G may be generated in the crust on the eve of coalescence from low-order gravity ($g$-) mode resonances (see also Ref.~\cite{rs25}).
While it was shown that chemical and thermal diffusion should not suppress the instability if the star is rotating sufficiently fast ($\nu = \Omega/2\pi \gtrsim 20 \text{ Hz}$), elasticity was ignored.
By revisiting the classical MRI analysis with elasticity, one goal of this paper is to assess to what extent the instability is quenched in a neutron-star crust.
At the risk of spoiling the reader, the main conclusion is that it will be \emph{unless} the star is rotating very fast ($\nu \gtrsim 300$~Hz).
Neutron stars with spins of this magnitude are unlikely to participate in a merger, though some evolutionary channels involving recycling could produce them \cite{ff12}. 

The fact that only a small fraction\footnote{It has been suggested that the mass-gap ($2.6 M_{\odot}$) secondary in the merger event GW190814 was a neutron star with spin period $P \sim 1$~ms (see, e.g., Ref.~\cite{tso20}). If this is the case, rapid rotation may thus not be so rare.} \cite{zhu18} of pre-merger neutron stars may be rapidly rotating is not altogether discouraging though.
Indeed, a $\sim$~few percent of merger-driven gamma-ray bursts (GRBs) exhibit what are known as \emph{precursor flares}: short flashes of high-energy emission up to $\sim 20$~seconds prior to the main event (see Ref.~\cite{skk24} for a review).
If these emissions are to be produced \emph{before} merger, strong magnetic fields are required to match the observed luminosities \cite{tsan12,tsan13,most20,most22}.
Preserving a crustal field of strength $\gtrsim 10^{13}$~G in a $\sim$~Gyr old neutron star appears impossible however from a magnetothermal perspective \cite{pv19}, and thus it could simply be the case that only rapidly-rotating stars are responsible for launching precursor flares.
Alternatively, as viscous heating during inspiral progressively melts the outer layers of the crust \cite{Arras18,pan20,ghosh24}, the resulting ocean may be deep enough that the familiar MRI analysis can be reliably applied.
Gravitational wave (GW) \cite{east19,rezz26,karak26} and kilonovae \cite{ross24} signals will also be influenced by spin, providing an independent channel to test the framework. 

This work is organised as follows. 
In Section~\ref{sec:setup} we introduce the elastic-MHD equations of motion, which are then linearised about a background shear flow (Sec.~\ref{sec:perts}) and projected into plane waves (Sec.~\ref{sec:plane}), to set up the problem.
Dispersion relations are derived in Section~\ref{sec:uniform} for uniform solids, which are then analysed to deduce the existence (or otherwise) of growing modes which represent the MRI that can amplify a weak, preexisting magnetic field (Sec.~\ref{sec:satampuniform}); some remarks on the spatially-varying case are given in Sec.~\ref{sec:varying}.
The scheme is then applied to the case of neutron-star crusts in Section~\ref{sec:nscrust}, with the implications for precursor flares explored in Section~\ref{sec:finalwords}.
Some closing discussion regarding future directions and astrophysical caveats are given in Section~\ref{sec:discussion}.

\section{Equations of motion and setup} \label{sec:setup}

We consider an inviscid fluid rotating differentially with velocity $\bv$ in a cavity hosting an ambient magnetic field, $\boldsymbol{B}$. 
Working within a Newtonian framework, the equations of momentum balance read  \cite{fuller14,bera20}
\begin{equation} \label{eq:momentum}
0 = \rho \frac{D \boldsymbol{v}}{D t} + \bnabla p - \bnabla \cdot \bt + \rho \bnabla \Phi - \frac{1}{4\pi} (\bnabla \times \boldsymbol{B}) \times \boldsymbol{B},
\end{equation}
for advective derivative $D = \partial_{t} + \bv \cdot \bnabla $, mass-density $\rho$, pressure $p$, and gravitational potential $\Phi$. 
The main new piece of physics we consider in this paper, going beyond previous studies, is that we allow the cavity medium to be \emph{elastic}. 
For this purpose, we follow the notation of \citet{fuller14} and consider purely adiabatic and elastic oscillations.
Elastic effects are thus captured by the anisotropic stress tensor, $\bt$, with components\footnote{Assuming Hooke's law and ignoring higher-order Lam{\'e} coefficients; cf. \citet{sot24} and references therein.}
\begin{equation} \label{eq:tauel}
\tau_{\rm el}^{ij} = \mu\left[ \left( \frac{\partial \xi^i}{\partial x_j} + \frac{\partial \xi^j}{\partial x_i} \right)  - \frac{2}{3} \left( \bnabla \cdot \bxi \right)  g^{ij} \right],
\end{equation}
where $\mu$ represents the shear modulus, $\boldsymbol{\xi}$ is the Lagrangian displacement field, and $\boldsymbol{g}$ is the Euclidean metric. 
In the electromagnetic sector, we have the induction equation,
\begin{equation} \label{eq:induction}
0 = \frac{\partial \boldsymbol{B}}{\partial t} - \bnabla \times (\boldsymbol{v} \times \boldsymbol{B}),
\end{equation}
where we ignore dissipative terms related to magnetic diffusion. 
Equations \eqref{eq:momentum} and \eqref{eq:induction}, when supplemented by the continuity equation,
\begin{equation} \label{eq:continuity}
0 =    \frac{\partial \rho}{\partial t} + \bnabla \cdot \left( \rho \bv\right),
\end{equation}
the solenoidal condition on the magnetic field, $\bnabla \cdot \bb = 0$, the Poisson equation for $\Phi$, and an equation of state (EOS), fully define the system of interest subject to appropriate boundary conditions.

\subsection{Linear perturbations} \label{sec:perts}

Although one could write down the elastic-MHD perturbation equations for a general background, we make a number of simplifications to investigate the MRI in a restricted context. 
In particular, we consider a simple background rotation of the form $\boldsymbol{\Omega} = \Omega(r) \hat{\boldsymbol{z}}$ and a constant, vertical magnetic field $\boldsymbol{B}= B_z \hat{\boldsymbol{z}}$ in locally-Cartesian coordinates ($x,y,z$). 
We work within a rotating frame, and introduce Eulerian variations ($\delta$) of the equilibrium vectorial quantities through
\begin{equation}
\boldsymbol{v} \to \boldsymbol{v} + \delta \boldsymbol{v}, \quad \boldsymbol{B} \to \boldsymbol{B} + \delta \boldsymbol{B},
\end{equation}
together with the scalar ones in the obvious way.
Perturbing each variable in this manner, we linearise the equations of motion about the background to arrive at
\begin{equation} \label{eq:linmom}
\begin{aligned}
0 =& \rho \left(\frac{\partial \delta \bv}{\partial t} +  2 \boldsymbol{\Omega} \times \delta \bv \right) + \delta \rho \frac{D \boldsymbol{v}}{Dt} + \bnabla \delta p \\
&- \bnabla \cdot \delta\bt - \frac{1}{4\pi} (\bnabla \times \delta \boldsymbol{B}) \times \boldsymbol{B},
\end{aligned}
\end{equation}
\begin{equation} \label{eq:lininduction}
    0 =\frac{\partial \delta \boldsymbol{B}}{\partial t} - \bnabla \times (\boldsymbol{\delta v} \times \boldsymbol{B} + \bv \times \delta \bb),
\end{equation}
and
\begin{equation} \label{eq:lincty}
    0 =    \frac{\partial \delta\rho}{\partial t} + \bnabla \cdot \left( \delta\rho \bv + \rho \delta \bv\right),
\end{equation}
where we ignore gravitational perturbations ($\delta \Phi = 0$) by adopting the Cowling approximation\footnote{We note that in the classical picture of MRI in accretion disks, perturbations of the gravitational potential are neglected because the fluid is assumed to be non–self-gravitating. In the present context, this constitutes an additional, and perhaps strong, assumption.}.
Importantly, we further assume that the background is in a relaxed state such that $\bxi$ is strictly an $\mathcal{O}(\delta)$ quantity and thus $\delta \bt$ is, through a minor abuse of notation, also described by equation \eqref{eq:tauel}. In general, the Lagrangian variation ($\Delta$) of a (tensorial) quantity $\boldsymbol{X}$ is related to Eulerian perturbations through \cite{lo67,fs78}
\begin{equation} \label{eq:generalvar}
    \Delta \boldsymbol{X} = \delta \boldsymbol{X} + \mathcal{L}_{\xi} \boldsymbol{X},
\end{equation}
for Lie derivative $\boldsymbol{\mathcal{L}}$. 
For stagnant fluids, we have the simple relation $\Delta \bv = \delta \bv = \partial \bxi / \partial t$ by definition. 
In cases with non-trivial background flows, however, the more complicated relation,
\begin{equation} \label{eq:xidefn}
\delta \bv = \frac{\partial \bxi}{\partial t} + \left( \bv \cdot \bnabla \right) \bxi - \left( \bxi \cdot \bnabla \right) \bv,
\end{equation}
must be used to relate velocity perturbations to displacements (see equation 5 in \citet{fs78}). In terms of the Lagrangian displacement, we note that the linearised continuity equation \eqref{eq:lincty} reduces to an algebraic one \cite{lyn67}, viz.
\begin{equation} \label{eq:ctysimpd}
    \delta \rho = - \bxi \cdot \bnabla \rho - \rho \bnabla \cdot \bxi.
\end{equation}

Within the local Cartesian frame introduced above, we adopt a linear background shear flow of the form 
\begin{equation} \label{eq:shearprofile}
    \bv =  S x \hat{\boldsymbol{y}},
\end{equation}
in accord with the usual MRI prescription \cite{bal91,bal98}.
Here, $S = {d \Omega }/{d \left(\log r\right)}$ denotes the shear rate. 
For this flow, equation \eqref{eq:xidefn} thus gives us the components of the displacement vector through
\begin{equation} \label{eq:xix1}
    \frac {\partial \xi_{x}}{\partial t} = \delta v_{x},
\end{equation}
\begin{equation} \label{eq:xiy1}
        \frac {\partial \xi_{y}}{\partial t} = \delta v_{y} + S \xi_{x},
\end{equation}
and
\begin{equation} \label{eq:xiz1}
        \frac {\partial \xi_{z}}{\partial t} = \delta v_{z} .
\end{equation}
Even at this level, it is clear that solving \eqref{eq:linmom}--\eqref{eq:lincty} is challenging and requires numerical treatment in general for any given background density, pressure, and so on. 

The context of the problem we are considering is that of fluid oscillations within a radial cavity, $R_{\rm cc} \leq r \leq \Rstar$. 
For a neutron star, these would represent the size of the crust, with $R_{\rm cc}$ denoting the crust-core interface and $\Rstar$ being the radius of the stellar surface. 
Boundary conditions apply at these two edges (e.g., that the Lagrangian variation of the pressure, $\Delta p$, should vanish) though the details are unimportant in the simplified treatment we apply with respect to a locally plane-parallel expansion.

\subsection{Plane-parallel geometry and WKB expansions} \label{sec:plane}

In accord with the simple, albeit standard, treatment of the MRI, we consider a localised fluid element within a plane-parallel geometry, where most background quantities can be treated as effectively constant.
In this case, perturbations can be described via plane waves to examine the frequencies and thus propagation of MHD modes. 
This is formally handled through Wentzel–Kramers–Brillouin (WKB) expansions, where we effectively assume that the wave properties (e.g., frequency) vary slowly enough in space such that perturbations can be approximated through local plane-waves whose parameters depend on position. 
This is reasonable provided wavelengths of the perturbations are much smaller than the characteristic scales of variation of the background; formally, perturbations are handled through (see, e.g., Refs.~\cite{cel24,cel25})
\begin{equation}
\begin{aligned}
    \boldsymbol{X} &\to \boldsymbol{X} + \delta \boldsymbol{X} \\
    &=\boldsymbol{X} + \bar{\delta}\left(\sum_{q=0}^{\infty}\varepsilon^q \boldsymbol{X}_q\right) e^{i\theta(\boldsymbol{x})/\varepsilon} + \mathcal{O}(\bar\delta^2) \;,
    \end{aligned}
\end{equation}
{for some phase $\theta$ and book-keeping parameters $\bar{\delta}$ (small-amplitude expansion) and $\varepsilon$ (short-wavelength expansion). 
The latter quantify the relative size of the background against perturbations, with $\varepsilon = \lambda/L \ll 1$ being the ratio between $\lambda$ -- the typical wavelength of a given mode -- and $L$, being the typical length-scale over which the amplitude of the mode varies.
The phase, $\theta$, encapsulates the kinematics in the sense that its gradient defines the local wavevector, $\boldsymbol{k} = \nabla \theta$.
The integer $q$ controls the asymptotic series: $q=0$ terms give rise to the leading-order dispersion relation, $q=1$ relates to transport and how the wave amplitude evolves in time, and higher-order terms account for mode-medium couplings and nonlinear effects.}
Within this framework we can also account for a non-homogeneous background flow, provided it varies slowly in space, e.g., it is sheared with shear rate $S=S(\varepsilon x)$.
The dispersion relation in the presence of such a background can then be obtained by formally substituting this expansion and retaining only the lowest-order terms $\mathcal{O}(\bar\delta, \varepsilon^0)$.

In the simple scheme considered here, we work with purely vertical modes with wavenumber $k$ and assume all perturbed variables grow or are damped at the same rate, $\sigma$. 
In symbols, this means we apply the ans{\"a}tze
\begin{equation} \label{eq:verticalansatz}
\delta \boldsymbol{v}, \delta \boldsymbol{B}, \boldsymbol{\xi}, p \propto e^{i k z + \sigma t},
\end{equation}
with the perturbed density obtained directly from equation \eqref{eq:ctysimpd}.
The decomposition implied by expression \eqref{eq:verticalansatz} effectively reduces the problem to an \emph{algebraic} one which can be tackled analytically.

\section{Uniform solids} \label{sec:uniform}

In the context of our problem of the MRI within elastic media, we note that it is only the momentum-balance equation \eqref{eq:linmom} which features the relevant modulii: the induction \eqref{eq:lininduction} and continuity \eqref{eq:ctysimpd} equations are identical to the liquid case. 
The solutions are thus the classical ones. 
The magnetic-field amplitudes read
\begin{equation}
\delta B_{x} = \frac{i B_{z} k \delta v_{x}}{ \sigma }, \delta B_{y} = \frac{i B_{z} k \left(S \delta v_{x}+\sigma  \delta v_{y}\right)}{\sigma^2},
\end{equation}
with $\delta B_{z} = 0$ from the solenoidal condition, $\bnabla \cdot \delta \bb = 0$.

Suppose for now that $\mu$ is constant throughout the cavity for simplicity,
Using the displacement \eqref{eq:xix1}--\eqref{eq:xiz1} we find, from the $z$-component of equation \eqref{eq:linmom}, the amplitude of the pressure perturbation,
\begin{equation} \label{eq:pressurepert}
    \delta p = i \frac{4\mu k^2 + 3\rho \sigma^2}{3\sigma k}\delta v_z.
\end{equation}
Notably, we see that $\delta p$ depends linearly on $\mu$. 
This is physically expected, and is simply echoing the fact that allowing for the elastic medium to be sheared necessarily implies restorative pressure disturbances. 
In general, one must impose an isentropic or other condition to relate the perturbed pressure to the density -- and hence $\delta v_{z}$ -- to close the system.
For the case under consideration though, $\delta v_{z}$ does not influence the dynamics and we can ignore such an imposition without loss of generality.
By writing out the remaining two momentum equations and building a coefficients matrix for $\delta v_{x}$ and $\delta v_{y}$, we take its determinant to obtain the dispersion relation,
\begin{equation} \label{eq:cstsheardisp}
\begin{aligned}
0=& \,   \sigma^4 + \sigma^2\frac{ k^2 \left( B_z^2 + 4 \pi \mu \right) + 4 \pi \rho \Omega (S + 2\Omega)}{2 \pi \rho} \\
    &+ \frac{k^4 \left( B_z^2 + 4 \pi \mu \right)^2 + 8 \pi k^2 \rho \left(B_{z}^2 + 4 \pi \mu \right) S \Omega}{16 \pi^2 \rho^2}.
\end{aligned}
\end{equation}
This equation, much like the liquid case, is quadratic in $\sigma^2$ and can be solved with ease. 
We see the key impact of elasticity is to shift the magnetic terms by $4 \pi \mu$. 

Equation \eqref{eq:cstsheardisp} can be cast in a more symmetrical form by introducing the effective (angular) Alfv{\'e}n, $\omega_{\rm A}^2 = B_z^2 k^2/4 \pi \rho$, and shear, $\omega_{\rm S}^2 = \mu k^2 /\rho$, velocities respectively. 
Doing so returns
\begin{equation} \label{eq:cstsheardisp2S}
\begin{aligned}
0=&\,    \sigma^4 + 2 \sigma^2 \left[ \Omega \left(S + 2 \Omega \right) + \omega_{\rm A}^2 + \omega_{\rm S}^2 \right] \\
    &+ \left( \omega_{\rm A}^2 + \omega_{\rm S}^2 \right) \left( 2 S \Omega + \omega_{\rm A}^2 + \omega_{\rm S}^2 \right).
\end{aligned}
\end{equation}
Within the quadratic and constant (in $\sigma$) terms, we see that it is the combination $\omega_{\rm A}^2 + \omega_{\rm S}^2$ which appears; in fact, the only difference with the liquid case is that $\omega_{\rm A}^2 \to \omega_{\rm A}^2 + \omega_{\rm S}^2$. 
This is because the magnetic and elastic oscillation modes effectively merge into a family of hybrid Alfv{\'e}n-shear modes (see, e.g., Refs.~\cite{piro05,ck11}). 
The fact that it is the sum $\omega_{\rm A}^2 + \omega_{\rm S}^2$ that appears demonstrates that elasticity is strictly restorative and thus reduces the saturation amplitude for a magnetic field growing through any would-be instability. 
Intuitively, if $\omega_{\rm A}^2 + \omega_{\rm S}^2$ were equal to a positive constant then $\omega_{\rm A}^2$ must be lower in elastic media relative to liquids. 
This is important since, by the Routh-Hurwitz theorem, it is the positivity of the $\sigma^2$ and $\sigma^0$ coefficients which are responsible for determining whether the system is (un)stable \cite{anag91}. 

It is worth commenting that from these simple relations we can see that the growth rate of the instability matches that of the liquid case, though the wavelength is shifted through the shear modulus.
The story changes slightly if the shear modulus varies in space (see Sec.~\ref{sec:varying}), but the main conclusion is that the saturation value of the MRI changes but not the amplification timescale, as we explore next.

\subsection{Magnetic saturation} \label{sec:satampuniform}

To find the saturation field corresponding to the most unstable mode, we first differentiate expression \eqref{eq:cstsheardisp2S} with respect to $k$ to determine the roots of $0 = \partial \sigma / \partial k$. 
From the second derivative test, these roots can be used to identify the modes with the highest growth rate -- as relevant for the MRI. Carrying this out returns the condition
\begin{equation} \label{eq:dsigmadkS}
\begin{aligned}
 0&=   \sigma^2 + S \Omega + \oA^2 + \oS^2.
\end{aligned}
\end{equation}
Isolating $\sigma^2$ and resubstitution into \eqref{eq:cstsheardisp2S} in turns yields the remarkably simple relationship
\begin{equation} \label{eq:genMRI1S}
\begin{aligned}
    0 &= S^2 + 4 S \Omega + 4 \left(\oA^2 + \oS^2 \right).
\end{aligned}
\end{equation}
Equation \eqref{eq:genMRI1S} is effectively an \emph{elastic generalisation} of the (basic) MRI formula for the saturation field\footnote{More precisely, this gives an estimate for the magnetic field at saturation, based on extrapolating linear theory results until saturation. For a more detailed analysis see, e.g., Ref.~\cite{MiravetPessah25}.}; it describes the minimum magnetic tension required to stabilise the system. 
In terms of our primitive variables and wavelength $\lambda = 2 \pi/k$, it can be rewritten as
\begin{equation} \label{eq:primitiveexpression}
    \frac{ 4 \pi \left(B_{z}^2+4\pi\mu\right)}{\lambda^2 \rho} = - \left( 1 + \frac{\kappa^2}{4 \Omega^2} \right) \frac{d \Omega^2}{d \log r} ,
\end{equation}
where $\kappa^2 = r^{-3} \partial_{r} \left( r^4 \Omega^2\right)$ is the square of the epicyclic frequency. 
It is clear from equation \eqref{eq:primitiveexpression} that we recover the standard liquid limit when $\mu \to 0$ (see, e.g., Ref.~\cite{bal98}). 

\begin{figure}
\centering
  \includegraphics[width=0.487\textwidth]{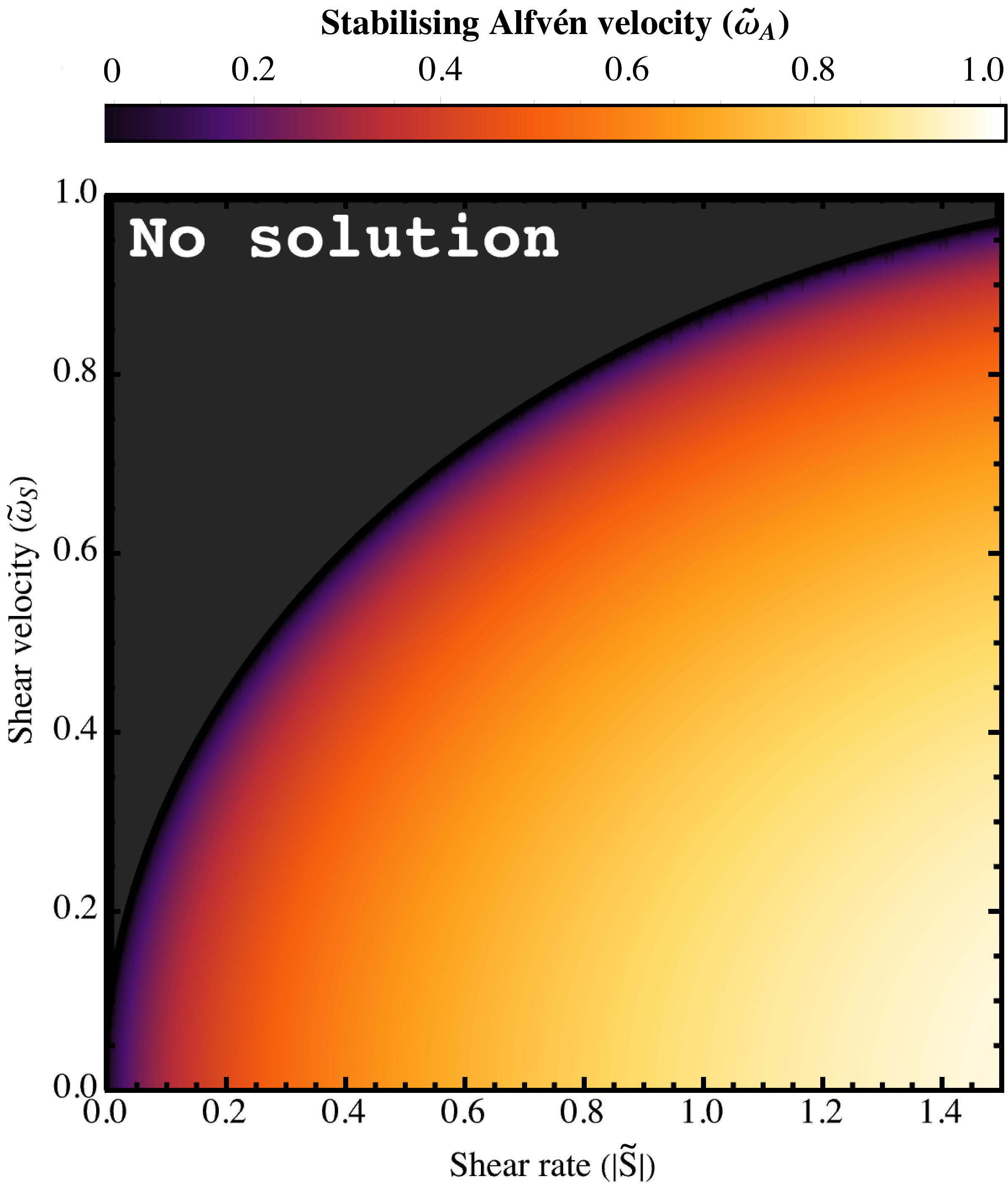}
  \caption{Stabilising Alfv{\'e}n velocity for constant shear modulus (i.e., the solution to equation \ref{eq:dimensionlessshear}) as a function of dimensionless shear rate, $\tilde{S}$, and shear velocity, $\tilde{\omega}_{\rm S}$. The dark, shaded region corresponds to cases where no solution exists: elasticity is sufficient to prevent instability. Brighter shades indicate stronger magnetic fields are required for stabilisation.
  }
  \label{fig:omegas_simplified}
\end{figure}

It is convenient to introduce dimensionless variables, $\tilde{S} = S/\Omega$ and $\tilde{\omega}_{\rm S,A} = \omega_{\rm S,A}/\Omega$, so that expression \eqref{eq:genMRI1S} can be re-arranged to give
\begin{equation} \label{eq:dimensionlessshear}
    \tilde{\omega}_{\rm A}^2 = -\frac{1}{4}\left( \tilde{S}^2 + 4 \tilde{S} + 4 \tilde{\omega}_{\rm S}^2\right).
\end{equation}
Figure~\ref{fig:omegas_simplified} depicts the relationship \eqref{eq:dimensionlessshear}, which describes the stabilising Alfv{\'e}n velocity as a function of shear rate and shear velocity. 
The stabilising influence of elasticity is clear: if $\omega_{\rm S}$ is sufficiently large for a given $|S|$, the system is stable and no magnetic growth is expected. 
We see that even cavities with significant, normalised shear rates of $|\tilde{S}| = 0.5$ are unconditionally stable for $\tilde{\omega}_{\rm S} \geq \sqrt{7}/4 \approx 0.66$. 
For smaller shear modulii, the value of $\tilde{\omega}_{\rm A} \propto B_{z}$ required to halt the instability increases effectively linearly. 
As expected, the greatest magnetic field amplifications are anticipated for the largest $|S|$ and vanishing $\omega_{\rm S}$ (i.e., values in the bottom-right corner). 
With respect to physical units, the system tends to be more unstable for large values of $\Omega$: larger $\omega_{\rm S}$ are required, for any given $|S|$, to prevent instability. 
From equation~\eqref{eq:genMRI1S}, it is easy to see that, for a given $|S|$, the required $\omega_{\rm S}$ increases by a factor $\sqrt{2}$ when doubling $\Omega$. 
This means that the MRI overcomes the stabilising effects of elasticity more easily when the system is rotating rapidly.

\subsection{A glimpse at the varying modulus case} \label{sec:varying}

We here consider the case in which the shear modulus is not constant but varies in space. This situation can be treated within the WKB framework outlined in Sec.~\ref{sec:plane}. 

In this case, it is convenient to introduce the quantity
\begin{equation} \label{eq:gammax}
    \gamma_{x}^2 = k(\partial_x \mu)/\rho,
\end{equation}
which sets the characteristic frequency scale associated with radial variations of the shear modulus (and analogously for $\gamma_z$). 

Writing out the elastic stress tensor, we find
\begin{equation}
    \partial_j \tau^{jl}_{el} = \begin{cases}
        -\rho \omega_S^2 \xi_x + i \rho\gamma_z^2 \xi_x -i\frac{2}{3}\rho\gamma_x^2\xi_z \;\quad & l=x, \\ 
        -\rho \omega_S^2 \xi_y + i\rho\gamma_z^2 \xi_y \;\qquad\qquad \qquad  &l=y, \\
        -\frac{4}{3}\rho\omega_S^2\xi_z + i\rho\gamma_x^2 \xi_x + \frac{4}{3}i \gamma_z^2 \xi_z \;\quad &l=z.
    \end{cases}
\end{equation}
Proceeding as in the constant-modulus case, we first obtain the modified pressure perturbation,
\begin{equation}
    \delta p = i \frac{4\mu k^2 + 3\rho \sigma^2}{3\sigma k}\delta v_z + \frac{\rho}{3k}\left(4\gamma_z^2\delta v_z - \gamma_x^2\delta v_x \right).
\end{equation}
From this expression, we observe that: 
(i) the first block of terms is identical to the constant-modulus case and represents elastic (sound-like) waves in the solid; 
(ii) the second term corresponds to a gradient-induced coupling term for $z$-polarized waves; and 
(iii) the last term corresponds to a gradient-induced coupling term associated with $x$-polarized waves.

As before, one can compute the modifications to the $x$ and $y$ components of the Euler equations to derive the dispersion relation for $\sigma$. 
{Without further simplifications though}, the additional terms couple all three Euler equations and, generically, the system cannot be solved without specifying an EOS or imposing, for instance, incompressibility. 
These additional terms reflect the dispersive effect of spatial gradients in the shear modulus, breaking the decoupling between polarization components at linear order.

{Nonetheless, analytical progress can be achieved in two simplified settings: (i) by assuming that the shear modulus varies only in the vertical direction, such that $\gamma_x = 0$; or (ii) allowing both $\gamma_x,\,\gamma_z \neq 0$ but instead imposing incompressible perturbations, as commonly done in studies of the MRI. 
Both assumptions lead to the same decoupling observed in the constant shear modulus case---so that we can derive the dispersion relation from the coefficients matrix for $\delta v_x,\,\delta v_y$---and, remarkably, both yield the same dispersion relation
\begin{equation} \label{eq:newdisp}
\begin{aligned}
    0=\,& \sigma^4 + 2\sigma^2\left[\Omega(S+2\Omega) + \omega_T^2 - i\gamma_z^2\right] \\
    &- 2i \Omega \sigma \gamma_z^2 +\left(\omega_T^4 + 2\Omega S\omega_T^2 - 2i \gamma_z^2\omega_T^2 - \gamma_z^4 \right)\;,
\end{aligned}
\end{equation}
where $\omega_{\rm T}^2 = \omega_{\rm A}^2 + \omega_{\rm S}^2$. Note that equation \eqref{eq:newdisp} is no longer an even quartic polynomial in $\sigma$ (a non-vanishing linear term appears) and the coefficients multiplying the even powers of $\sigma$ acquire imaginary contributions.}

While a detailed analysis of this dispersion relation generally requires numerical treatment and lies beyond the scope of this work, some analytical insight can still be gained through a perturbative approach. 
Specifically, we treat the terms arising from the spatial variation of the shear modulus as small, introducing a book-keeping parameter $\epsilon$ (not to be confused with the WKB expansion parameter) to write $-i\gamma_z^2 \to -i\epsilon\gamma_z^2$. We then write $\sigma = \sigma_0 + \epsilon\sigma_1 + \mathcal{O}(\epsilon^2)$, so that
\begin{align}
    \sigma^2 &= \sigma_0^2 + 2\epsilon\sigma_0\sigma_1  + \mathcal{O}(\epsilon^2), \\
    \sigma^4 &= \sigma_0^4 + 4\epsilon  \sigma_0^3\sigma_1 + \mathcal{O}(\epsilon^2) .
\end{align}
Substituting the above into the dispersion relation, we find that, at $\mathcal{O}(\epsilon^0)$, one recovers the constant-modulus solution as expected. At first order [$\mathcal{O}(\epsilon)$], we have instead
\begin{equation}
    \sigma_1 = i\gamma_z^2 \frac{2\sigma_0^2 + 2\Omega\sigma_0+2\omega_T^2}{4\sigma_0^3 + 4\sigma_0 (\omega_T^2 + \Omega(S+ 2\Omega)}.
\end{equation}
Specializing to the fastest-growing MRI mode, which in the elastic case also has growth rate $\sigma_0 = -S/2$, we obtain
\begin{equation} \label{eq:sigma1}
    \sigma_1 = -i \gamma_z^2 \frac{S^2 - 2\Omega S + 4\omega_T^2}{S^3 + 4S(\omega_T^2 + \Omega(S+2\Omega))}\;.
\end{equation}

Most importantly, we find that, in the presence of a spatially varying shear modulus, and provided that its gradients are sufficiently small, the effect is simply to introduce an imaginary correction to the growth rate, and therefore a real component to the frequency. 
In other words, modes that were purely growing or decaying acquire a propagating character. {Note that, at least in the neutron-star context, we also anticipate $\partial_z \mu <0$ as the modulus decreases as one moves towards the surface.
This implies that the waves will propagate towards lower density regions [as $\text{Im}(\sigma_1)  <0$] such that the system will naturally evacuate magnetic and kinetic energy towards the outer edges of the cavity. 
}

Crucially, in this regime the stability properties and saturation estimates remain essentially unchanged from the uniform-modulus case. 
In the context of neutron-star crusts, as we explore more thoroughly in Sec.~\ref{sec:hydrostatic}, a handy numerical estimate from equation \eqref{eq:sigma1} is $|\sigma_{1}| / |\sigma_0| \sim 0.1$ for shear rates $|S| \sim 600 \text{ s}^{-1}$.
This implies that:
(i) since the ratio is at most $\sim 10\%$, the perturbative analysis is internally consistent, and 
(ii) results obtained within the simplified scheme of a uniform modulus are largely applicable even to more realistic cases except near the crust-ocean interface where $\mu$ varies strongly.
We henceforth focus on the constant modulus case. 

\section{Application to neutron-star crusts in binary mergers} \label{sec:nscrust}

Having generalised the equations of motion describing activation criteria for the MRI to elastic media, we now turn to analysing the case of a neutron-star crust.
In a mature neutron star, the crustal cavity begins when the density drops below a value of $\rho_{\rm cc} \approx 1.4 \times 10^{14} \text{ g cm}^{-3}$, which typically occurs at a radius of $r = R_{\rm cc} \approx 0.9 R$ in realistic EOS.
Although the fluid variables span many orders of magnitude within such a domain, we can obtain some rough estimates for magnetic growth expectations by considering a constant density. 
This builds on the ideas presented in SKRK24, where elastic terms where ignored (see also Refs.~\cite{rs25,yong25}).
{To relate the mass and radius of the star, it is typically necessary to consider an EOS and solve the equations of hydrostatic equilibrium. 
To avoid such a complication, we simply fix $M = 1.4 M_{\odot}$ and $\Rstar = 12$~km to be roughly consistent with recent multimessenger constraints (see, e.g., figure 5 in Ref.~\cite{ofen24}) but we present formulae scaled by mass and radius where appropriate.
}

The problem explored in SKRK24 concerns differential rotations induced by the excitation of resonant oscillation modes within the final $\sim$~seconds of inspiral in a merger \cite{kuan21}. 
It was argued that, if elasticity can be ignored, the shear that results could excite the MRI when the mode phase is such that $d \Omega / dr <0$ and the eigenfunction hosts no nodes in the cavity.
In an effort to keep this work self-contained but concise, we begin by briefly reviewing the degree to which resonant modes introduce shear and relate that to our local variable, $S$, appearing within expression \eqref{eq:shearprofile}.

\subsection{Dynamical tides and Rossby numbers} \label{sec:dyntides}

Neutron stars are rich systems which host a plethora of oscillation modes. 
Roughly speaking, each fluid degree of freedom responds differently to perturbations and the set of responses defines the overall spectrum.
One prominent class involves buoyancy-restored ($g$-) modes, which generally exist when perturbations are non-isentropic and the star hosts compositional or temperature inhomogeneities \cite{finn87,rg92}.
These modes have received an amount of attention in the literature, as their frequencies lie in a range that is favourable to resonances with the orbital motion.

Particulars of the $g$-mode spectrum depend on the properties of the star
through the EOS, mass, and rate of beta re-equilibration \cite{ande19,kuan22,coun25}.
The $g$-modes most relevant in the context of tides are the $g_{1}$ and $g_{2}$ non-axisymmetric ($\ell = m =2$) modes, as these overlap the strongest with the tidal potential. 
In SKRK24, the general-relativistic pulsation equations were coupled to a high-order post-Newtonian scheme describing the orbital decay due to radiation-reaction and then solved numerically to deduce the amplitudes achieved by the modes at resonance. 
By extracting the azimuthal component of the velocity vector, SKRK24 showed that $g_{1}$ and $g_{2}$ resonances typically introduce \emph{Rossby numbers},
\begin{equation} \label{eq:rossby}
    \Ro \equiv \left|\frac{\Omega(\Rstar) - \Omega(R_{\rm cc})}{\Omega(R_{\rm cc})} \right|,
\end{equation}
of order $\sim 0.02$ and $\lesssim 0.1$, respectively. 
The exact value of $\Ro$ depends on the EOS and compactness, and could be larger by a factor of $\sim 2$ if the star is strongly stratified or massive.
Nonlinear effects tend to enhance the shear through tidal wave-wave interactions and parasitic instabilities, suggesting that the values quoted above are underestimates \cite{rs25}.
Although we do not model the system self-consistently in this paper, we fix a value of $\Ro = 0.1$ for concreteness.
To that degree, we essentially remain agnostic about which mode is responsible for introducing shear: although $g$-modes can introduce differential rotation as discussed above, other resonant modes may also drive the system towards comparable values of $\Ro \sim 0.1$ at phase maximum. 
For instance, interface \cite{tsan12,yong25}, inertial \cite{xu17,rs25}, and even acoustic \cite{suvo20,kk22} modes may introduce strong shears if the star hosts sharp phase transitions or is rotating rapidly.

In any case, from the definition of the shear, $S = d \Omega / d \log r$, we can use expression \eqref{eq:rossby} to estimate
\begin{equation} \label{eq:shearrossby}
\begin{aligned}
    |S| &\approx \left| \frac{\Delta \Omega}{\Delta \log r} \right|\\
    &= \Ro\frac{\Omega(\Rstar)}{\log(\Rstar) - \log (R_{\rm cc})} \\
    &\approx 600 \left( \frac{\Ro}{0.1} \right) \left( \frac{\nu}{100 \text{ Hz}} \right) \text{ s}^{-1}.
    \end{aligned}
\end{equation}
In what follows, we take expression \eqref{eq:shearrossby} as input for the saturation condition \eqref{eq:primitiveexpression}, setting $\lambda_{\rm max} \approx \Rstar - R_{\rm cc}$ for the maximum wavelength that can fit inside the crustal cavity.

\subsection{Shear modulus} \label{sec:hydrostatic}

In order to quantify the shear resistance of the crust, we introduce the Coulomb parameter,
\begin{equation} \label{eq:coul}
    \Gamma = \frac{Z^2 e^2}{a_{i} k_{B} T},
\end{equation}
where $n_{i}$ is the ion number density, $a_{i} = (4 \pi n_{i}/3)^{-1/3}$ is the (Wigner-Seitz) cell radius, $Z$ denotes the number of protons per ion, $e$ and $k_{B}$ are the elementary charge and Boltzmann constants, respectively, and $T$ is the crustal temperature. 
Given some $\Gamma$, we can appeal to calculations from the literature using microphysical, thermodynamic perturbation theory to estimate the shear modulus (see, e.g., Refs.~\cite{stroh91,hh08,baiko11}). 
We adopt the molecular-dynamics fit from \citet{hh08},
\begin{equation} \label{eq:shearmod}
    \mu \approx \mu_{0} \left( 0.1106  - \frac{28.7}{\Gamma^{1.3}} \right) \frac{n_i Z^2 e^2}{a_{i}},
\end{equation}
where we include the normalisation constant $\mu_{0}$ so that $\mu = 10^{30} \text{ dyn cm}^{-2}$ at the crust-core interface for $T \to 0$ so as to not artificially inflate the shear modulus due to the use of a simplified EOS (see below).
Importantly, since $\Gamma \propto T^{-1}$, $\mu$ is lower in hotter stars. 
With respect to dynamical tides which instigate viscous heating, this implies that in the later stages of inspiral the stars effectively host smaller shear modulii. 
As such, any would-be MRI is easier to activate closer to coalescence; in the limit of extremely high temperature, for example, the entire crust melts and $\mu$ effectively vanishes (see Sec.~\ref{sec:temperature}). 
In particular, $\Gamma \approx 175$ corresponds to the melting temperature and the onset of the ocean layer \cite{gud83,far93}. 
 
At the level of the local analysis considered in this paper, it is sensible to consider a toy model for $n_{e}$ to provide some order-of-magnitude estimates; more realistic profiles, using EOS data and spatially-varying modulii, will be considered in future work.
We take a polynomial profile for the electron number density of the form
\begin{equation} \label{eq:nerelation}
    n_{e}(r) = n_{e,0} \left(1 - \frac{r}{\Rstar}\right)^{\alpha},
\end{equation}
for some constants $\alpha$ and $n_{e,0}$. 
The total fluid density is therefore given by
\begin{equation} \label{eq:properdensity}
    \rho = n_{e} m_{e} + n_{i} m_{i} \approx n_{i} m_{i},
\end{equation}
where $m_{i} \approx A m_{p}$ for proton mass $m_{p}$.
We consider a catalysed crust composed of iron, and thereby fix the nuclear parameters to $A = 56$ and $Z = 26$. 
{In reality, $Z$ varies considerably throughout the crustal cavity, from $\gtrsim 40$ at the bottom of the outer crust to smaller values as one approaches the envelope \cite{cham08}. 
Given that it is the combination $n_i Z^2$ that predominantly controls expression \eqref{eq:shearmod} however, accounting for such complications would provide sub-leading corrections relative to the use of an already-simplified density profile \eqref{eq:nerelation}.
In other words, a depth-dependent $Z$ could be accounted for via $n_{i}$ in terms of the $\mu$ profile.
Similarly, as we have $\Gamma \propto n_i^{1/3} Z^2/T$, slightly different temperature models could similarly absorb any shifts associated with a varying $Z$ in terms of the coupling strength.
In any case,} the constant $n_{e,0}$ is chosen such that the density \eqref{eq:properdensity} matches the crust-core value, $\rho_{\rm cc} \approx 1.4 \times 10^{14} \text{ g cm}^{-3}$, at $r = R_{\rm cc} \approx 0.9 \Rstar$. 
This model matches reasonably well with the fit detailed by \citet{lg19} for the SLy4 crust for $\alpha \sim 3$ (see their expression 25).
Note that, from the expectation of charge neutrality, we have $n_{e} \approx Z n_{i}$ where $n_{e}$ is the electron number density.
Thus, with the exception of temperature, the constants determining $\Gamma$ -- and hence $\mu$ -- are fixed.

\begin{figure}
\centering
  \includegraphics[width=0.46\textwidth]{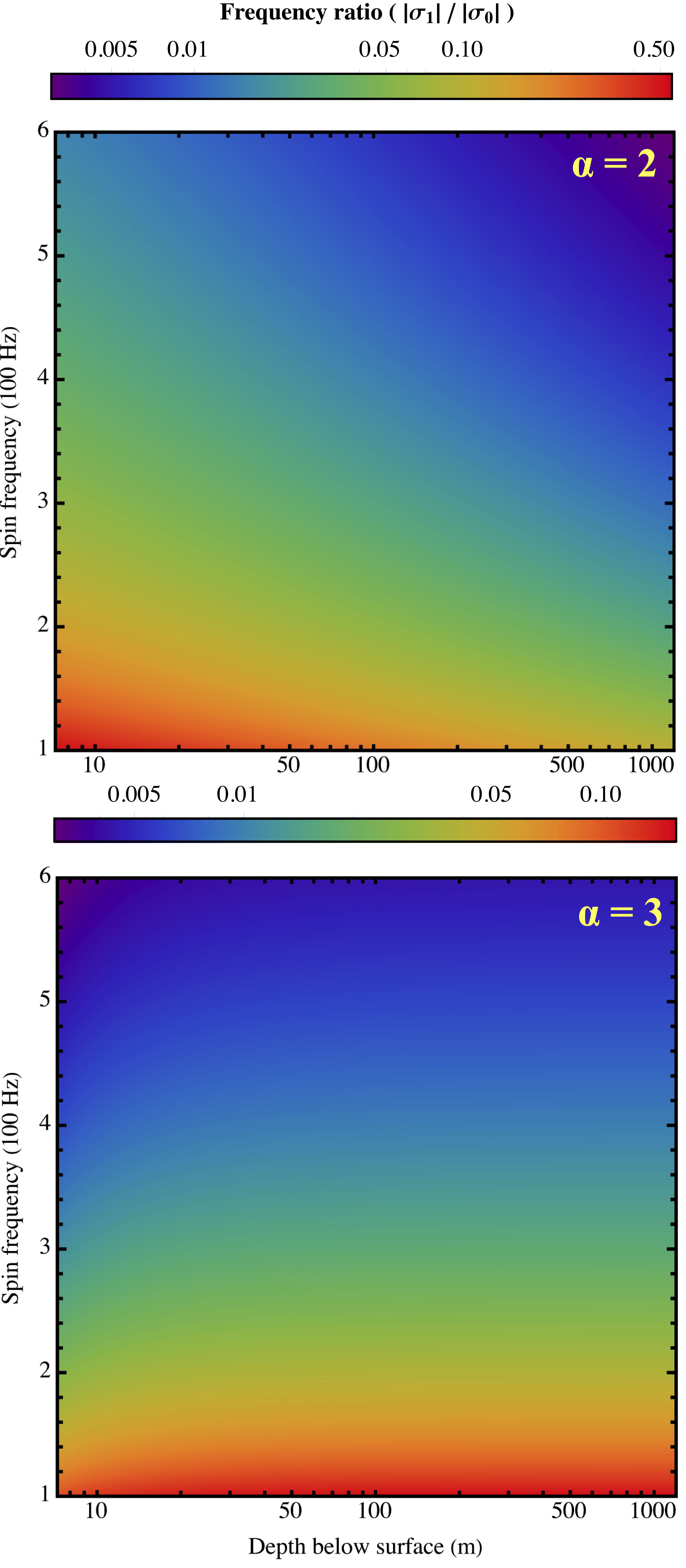}
  \caption{Absolute ratios between the zeroth- and first-order frequencies, $|\sigma_{1}| / |\sigma_0|$, as a function of depth and spin, as computed using equation \eqref{eq:sigma1} using the hydrostatic profiles specified in the main text for $\alpha =2$ (top) or $\alpha = 3$ (bottom).
  }
  \label{fig:ratio}
\end{figure}

Using these profiles, we can estimate the impact of spatial variations in the shear modulus, as in Sec.~\ref{sec:varying}. 
Figure~\ref{fig:ratio} displays the ratio $|\sigma_{1}| / |\sigma_0|$ from equation \eqref{eq:sigma1} for $T = 10^{8}$~K as a function of depth and spin frequency for shear rates corresponding to expression \eqref{eq:shearrossby}, where we approximate $\partial_{x} \approx \partial_r$ in expression \eqref{eq:gammax} for $\gamma_x$.
We consider two different electron number distributions, with either $\alpha = 2$ (top) or $\alpha = 3$ (bottom).
From expression \eqref{eq:gammax} we find $\gamma_x^2 \propto n_e'(r)/n_e(r)^{2/3}$ which varies little with depth in the $\alpha = 3$ case.
In this case we find, as quoted earlier, the frequency ratio is of order at most $\sim 10\%$ if $\nu \sim 100$~Hz and decreases monotonically with $\nu$ because $|\sigma_0| = S/2 \propto \nu$.
Even for $\alpha = 2$, the ratio is small except near the surface layers and low spins (though still is strictly less than unity).
These results therefore provide something of a retrodictive justification in our use of a constant $\mu$ for studying the MRI from mode resonances. 

\subsection{Temperature profile and viscous heating} \label{sec:temperature}

As modes are excited during the course of inspiral, viscosity leads to  heating.
This implies that the Coulomb parameter $\Gamma$ decreases as a function of time, which not only adjusts the shear modulus but also determines the ocean creep \cite{pan20} as the outer layers begin to melt.
However, the efficacy of viscous heating depends critically on the bulk viscosity which is largely unknown.

For example, \citet{lai94} argues that the \emph{non-resonant} fundamental ($f$-) mode in an aligned (or non-rotating) star in a binary will raise the stellar temperature to
\begin{equation}\label{eq:heat}
\begin{aligned}
    T(t) &= T_{i} + T_{H}(t) \\
    &\approx4\times10^{7}\left( \frac{3R}{a} \right)^{5/4} \text{ K},
    \end{aligned}
\end{equation}
where $a(t)$ is the orbital separation. 
{Note that expression \eqref{eq:heat} and others presented below are largely independent of the initial temperature, $T_{i}$, provided that it is much smaller than that induced by heating, $T_{H}(t)$, at the time of coalescence, $t_{c}$ [i.e., $T_{i} \ll T_{H}(t_c)$].}
Expression \eqref{eq:heat} likely provides an underestimate for $T_{H}$ however, as (i) other modes are excited beyond just the $f$-mode, and (ii) mode-splitting implies that rotating stars tend to accept resonances \emph{earlier} into inspiral (which leads to more heating; see Ref.~\cite{kuan23} for a discussion).
Moreover, if the core hosts a superfluid, the heat content is instead dominated by relativistic electrons and expression \eqref{eq:heat} may increase by factors of $\gtrsim 2$. 
If the core {composition is such that the direct Urca mechanism can activate \cite{sed24}}, the chemical heating driven by the dynamical tides could raise the temperature {to $T_{H}(t_{c}) \sim 2 \times 10^{8}$~K \cite{Arras18}. }
Stars containing strange matter, or which otherwise host huge bulk viscosities, will be even more susceptible to tidal heating; some authors estimate that even values {of $T_{H}(t_c)\sim 10^{10}$~K could} theoretically be reached \cite{ghosh24}. 
To account for such uncertainty, we adopt a simple parameterisation of the form
\begin{equation}\label{eq:tempparam}
    T(t) = T_i+4 T_{0} \left( \frac{3R}{a} \right)^{5/4},
\end{equation}
for some values $10^{7} \leq T_{0} /\text{K} \leq 10^{9}$; {the lower end reproduces the formula \eqref{eq:heat} from \citet{lai94}.}
Anticipating old neutron stars taking place in the merger, we assume an initial (volume-averaged) crustal temperature of $T_{i} = 10^{6}$~K in rough agreement with the temperature of the $\gtrsim$~Gyr old pulsar J0437--4715 \cite{karg04,gonz10} using the approximate crust-envelope temperature matching model from equation (8) in \citet{igo21}.

\begin{figure}
\centering
  \includegraphics[width=0.492\textwidth]{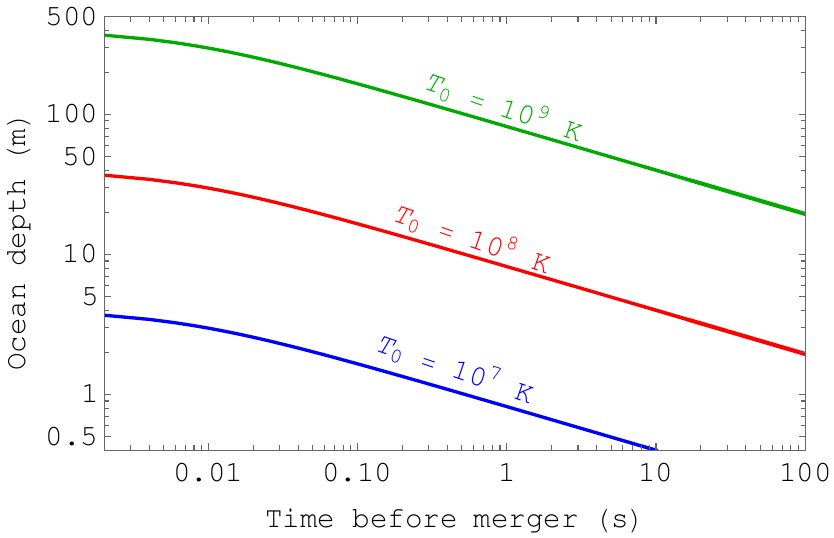}
  \caption{Effective ocean depth (where $\Gamma < 175$) as a function of time determined using toy models ($\alpha = 3$) describing orbital decay [Eq.~\eqref{eq:orbdecay}] and tidal heating [Eq.~\eqref{eq:tempparam}] for $T_{0} = 10^{9}$~K (green), $T_{0} = 10^{8}$~K (red), and $T_{0} = 10^{7}$~K (blue).
  }
  \label{fig:oceandepths}
\end{figure}

The time dependence in expression \eqref{eq:tempparam} enters implicitly through the orbital radius, $a$, which decays with time due to the emission of GWs (amongst other mechanisms \cite{skk24}). 
In particular, matching the GW luminosity with the time-derivative of the orbital energy implies \cite{mtw73}
\begin{equation}
\frac{da}{dt} = - \frac {64 G^3} {5 c^5} \frac {M^3 q \left( 1 + q \right)} {a(t)^3},
\end{equation}
for binary mass-ratio $q$. This equation has the well-known solution
\begin{equation} \label{eq:orbdecay}
a(t) = \frac {\left[ 81 c^5 R^4 - \tfrac {256} {5} G^3 M^3 q \left( 1 + q \right) \left( t - t_{\rm C} \right) \right]^{1/4} } {c^{5/4}},
\end{equation}
where $t_{\rm C}$ is the coalescence time, defined by the condition $a = 3 R$ if both stars have the same radius \cite{ho99}.

By substituting expression \eqref{eq:orbdecay} into \eqref{eq:tempparam} for some fiducial binary parameters ($q=1$), we can estimate the melting rate of the crust. 
Figure~\ref{fig:oceandepths} shows the depth corresponding to crystallization ($\Gamma = 175$) as a function of time relative to merger for several values of $T_{0}$. 
As expected, if the star heats more efficiently (larger $T_{0}$) we find that more of the crust has melted at any given pre-merger time. 
For example, for $T_{0} = 10^{9}$~K we see that the outermost $\sim 200$~m of the star may already be liquid $\sim 1$~s before merger where realistic, normal-fluid $g$-modes may become resonant \cite{kuan21,kuan22}. 
By contrast, if heating is inefficient and $T_{0} = 10^{7}$~K then even just before merger ($t - t_{\rm C} \sim 0.01$~s) we see that only several meters within the outermost layers will have melted.

\subsection{Magnetic field growth in the crust} \label{sec:applying}

\begin{figure*}
\centering
  \includegraphics[width=0.97\textwidth]{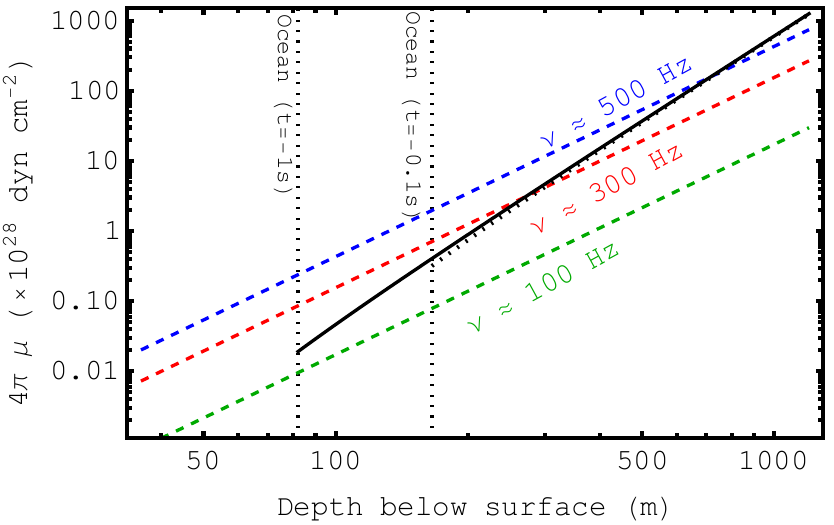}
  \caption{Comparison between $4 \pi \mu$ and the effective shear rates induced by resonant $g_{2}$ modes in an elastic crust (see text for details), for three representative background spin frequencies of $\nu \approx 100$~Hz (green, dashed), $\nu \approx 300$~Hz (red), and $\nu \approx 500$~Hz (blue). 
  Two shear modulii are considered, corresponding to either $\sim 1$~s prior to merger (solid line) or $\sim 0.1$~s prior (dotted line),
  as determined by the evolution of $\Gamma(t)$ for $T_{0} = 10^{9}$~K.
  The corresponding ocean depths at the relevant times are marked by vertical, dotted lines: $\mu \to 0$ at this boundary.
  Only when a given, dashed line lies above a solid one can the MRI activate, as per equation \eqref{eq:primitiveexpression}; the area bounding the two curves indicates the approximate saturation value of $B_{z}^2$ in units of $10^{28}$~G.
  }
  \label{fig:shearmod}
\end{figure*}

Using the toy profiles constructed above and values of $\Gamma(t)$ from expression \eqref{eq:coul}, we can estimate the shear modulus as a function of time.
Figure~\ref{fig:shearmod} compares $4 \pi \mu$ from expression \eqref{eq:shearmod} to the anticipated saturation amplitude for the magnetic field in the $\mu \to 0$ limit from expression \eqref{eq:genMRI1S} for several different base angular velocities.
Specifically, we show
\begin{equation} \label{eq:bsatbasic}
    B_{z,\rm max}^2 = -\frac{S \lambda^2 \rho \left( S + 4 \Omega \right)}{4 \pi},
    \end{equation}
where we adopt the shear rate \eqref{eq:shearrossby} and fix $T_{0} = 10^{9}$ to consider an optimistic scenario.
Such a comparison is motivated by the expectation that the MRI should \emph{not} operate if $4 \pi \mu$ exceeds the effective magnetic pressure from expression \eqref{eq:bsatbasic}.

As expected from equation \eqref{eq:shearrossby}, greater values of $\Omega = 2 \pi \nu$ induce more shear and thus a more powerful MRI.
On the `low' end with $\nu = 100$~Hz (green), we see that total suppression of the MRI is expected except perhaps in the very outer layers since $4 \pi \mu$ is sufficiently large that even for $B_{z} = 0$ the system is stable even for extreme heating ($T_{0} = 10^9$).
The implication is that only in the liquid ocean layers -- marked by the dotted vertical lines -- can the MRI operate in (comparatively) slow stars.
For faster stars, however, the shear is sufficiently strong that elasticity cannot halt the MRI in lower density crustal layers, up to depths of $\gtrsim$~300 meters for $\nu = 300$~Hz (red) or $\sim 1$~km for $\nu = 500$~Hz (blue). 
Note in particular that the units have been chosen such that the difference between a given dashed and solid line, if positive, indicates the value of $B_{z}^2$, in units of $10^{28} \text{ G}^2$, necessary to stabilise the system.
While stronger stratification or nonlinear effects \cite{rs25} could boost $\Ro$ to the point that lower values of $\Omega$ still permit instability in the elastic cavity, the overall conclusion is that only in very rapidly rotating stars do we expect an efficient MRI. 

\section{Astrophysical implications for precursor flares} \label{sec:finalwords}

Whether or not the MRI can activate in a premerger situation is critical for the ``energy crisis'' of \emph{precursor flares}. 
In particular, in a $\sim \text{few } \%$ of merger-driven GRBs, spikes of gamma-ray emissions have been observed anything up to about $\sim 20$~s prior to the main event \cite{skk24}. 
One promising (premerger) model to explain their spectra is crust yielding due to some mode reaching a high-enough amplitude at resonance to overwhelm the crystal lattice, thereby releasing an amount of magnetoelastic energy. 
A notable drawback of the model\footnote{Note that other models for premerger precursors (in the gamma-ray band), such as reconnections in a mutually-tangled magnetosphere \cite{most20,most22}, similarly require strong magnetic fields.} is that it is the magnetic field strength that controls the available energy. 
In particular, the maximum precursor luminosity from a failure event is of order \cite{tsan12,tsan13}
\begin{equation} \label{eq:maxlum}
L_{\rm prec} \sim 10^{47} \left(\frac{\delta v}{c}\right) \left( \frac {B_{\rm crust}} {10^{13}\, \mbox{G}}\right)^2 \left(\frac{\Rstar} {12\, \mbox{km}}\right)^{2} \mbox{erg\,s}^{-1},
\end{equation}
where $\delta v$ is the maximum velocity of the perturbation to the crust-anchored field lines.
As some precursors have reported luminosities exceeding $10^{49} \text{ erg s}^{-1}$ \cite{tsan23}, this implies that magnetar-like fields must somehow be preserved over cosmological timescales if the model is to apply.
This is difficult to reconcile with realistic magnetothermal evolutions regardless of the initial conditions \cite{pv19}.
It was this problem that led SKRK24 to investigate whether, if instead of being preserved, strong fields were generated on the eve of coalescence.
However, having arrived at the conclusion in the previous section that activating the MRI is challenging except for very high spins, the immediate temptation is to discard the possibility of pre-merger magnetic amplification unless $T_{0} \gg 10^{7}$ so that the ocean layer is wide enough to accommodate a sizeable MRI.

While spins of the order considered in Fig.~\ref{fig:shearmod} are unexpected in a typical binary merger, some evolutionary pathways may allow it.
For example, if one of the two neutron stars is recycled and has had its magnetic dipole field decay significantly over a cosmological timescale through magnetic burial or otherwise \cite{pm04,mukh17,suvm19}, much of the rotational kinetic energy could be retained up until merger. 
\citet{zhu18}, for instance, suggest that some realistic spin distribution priors for binary mergers indicate that values of $\chi_{\rm eff} \geq 0.05$ may occur in between $\sim$~2 and 6$\%$ of systems, where $\chi_{\rm eff}$ is the effective dimensionless spin of the binary. 
If one of the members is practically static within a system hosting a heavy and spinning neutron star, such a $\chi_{\rm eff}$ translates into a spin frequency of $200 \lesssim \nu / \text{Hz} \lesssim 300$ depending on the EOS (with softer/stiffer applying to the lower and upper ends, respectively).
As such, we could expect that -- at least in some rare instances -- the MRI may still play a r{\^o}le in the final seconds of inspiral.
Exactly how common this is depends on astrophysical evolutionary assumptions, an analysis of which lies beyond the scope of this article.
Nevertheless, numbers of a few percent are consistent with the observed precursor event rate, and we may thus posit that only those binaries with a rapidly rotating constituent can produce precursors.

The above hypothesis can be tested with future observations.
The orbital angular momentum for two equal stars just before merger (when $a \approx 3 R$) is of order 
\begin{align}
 L_{\rm orb} &\sim \frac{\sqrt{6}}{2} M \sqrt{G M R} \\
    &\approx 5 \times 10^{49} \left(\frac{M}{1.4 M_{\odot}} \right)^{3/2} \left(\frac{R}{12 \text{ km}}\right)^{1/2} \text{ g\,cm}^{2} \text{\,s}^{-1} \nonumber.
\end{align}
If we compare this with the spin angular momentum of a neutron star, $L_{\rm spin} \approx 2 \pi \nu I_{0},$ we find
\begin{equation} \label{eq:ratio}
\hspace{-0.3cm}    \frac{L_{\rm spin}}{L_{\rm orb}} \approx 0.06 \left( \frac{\nu}{300 \text{ Hz}} \right) \left(\frac{R}{12 \text{ km}}\right)^{3/2} \left(\frac{1.4 M_{\odot}}{M} \right)^{1/2},
\end{equation}
for moment of inertia $I_{0} \approx 0.38 M R^2$.
If angular momentum is approximately conserved, expression \eqref{eq:ratio} implies we expect the remnant to contain $\lesssim \mathcal{O}(10\%)$ more angular momentum in cases where the MRI may activate.

Since the rotational kinetic energy content of the remnant is likely to impact the efficacy of postmerger dynamo activity and jet production, we may expect a somewhat brighter GRB in cases with precursors (though cf. Refs.~\cite{kiu23,kiu25}).
We tentatively predict therefore a positive correlation between GRB luminosity and precursor emission.
Though no such luminosity trend for precursors relative to main events has been observed thus far to our knowledge, it is worth noting that two events, GRBs 211211A and 230307A, were both exceptionally bright and hosted precursors \cite{skk24}.
Additionally, while the fundamental ($f$-) mode couples much more strongly to the tides, it has a linear frequency of order $\sim 1$~kHz and thus will not become resonant prior to merger unless the stars are spinning rapidly with the correct orientation.
If indeed though a star rotates at a rate of $\nu \sim \mathcal{O}(300 \text{ Hz})$ in a merger, a resonance could trigger and lead to a substantial dephasing in the gravitational waveform that could be identified in future if enough cycles are observed \cite{kk22,kuan25}.
Additionally, the equilibrium tide becomes unstable to the emission of gravitational waves if the spin frequency of the primary exceeds the orbital frequency, instigating further dephasing  \cite{pnig19}.

\section{Discussion} \label{sec:discussion}

In this paper, we have reexamined the activation criteria for the MRI in a simplified, plane-parallel geometry following the analysis of Refs.~\cite{bal91,bal98} but in an \emph{elastic} medium. 
As elasticity is restorative, the additional terms appearing in the equations describing momentum balance \eqref{eq:momentum} can either lead to a weaker saturation field or shut off the MRI completely (Eq.~\ref{eq:primitiveexpression}). 
The result is physically intuitive: a solid lattice resists shear and, unless the restorative forces are completely overwhelmed, magnetic pressures are simply absorbed. 
This is quantified in Fig.~\ref{fig:shearmod} capturing our main results.
Indeed, only in rapidly rotating stars with $\nu \gtrsim 300$~Hz do we expect that realistic shears -- induced by resonant modes on the eve of coalescence -- are strong enough to overwhelm elastic restoration.
Even in this case, however, the MRI may still activate within the ocean where the shear modulus vanishes and the liquid criteria apply (Fig.~\ref{fig:oceandepths}).
If the temperature is high enough and the ocean layer deep enough (i.e., large enough $\lambda_{\rm max}$), sufficient flux may be produced to accommodate precursor luminosities (Eq.~\ref{eq:maxlum}). 

An important caveat to highlight with respect to our results is that we have restricted attention to plane-parallel perturbations: in cavities where the density, pressure, and shear modulus span several orders of magnitude (e.g., the crust), it would be more realistic to use spatially-varying profiles.
Upgrading the analysis in this way requires a thorough mode study however, where the full spectrum together with the relevant eigenfunctions are computed to identify the existence of Alfv{\'e}n-like modes undergoing exponential growth due to shear.
Such a task is left to future work, as realistic micro- and macrophysical inputs are needed which precludes analytic results (though see Sec.~\ref{sec:varying} for a glimpse).

Additionally, the local analysis carried out here makes no predictions for the \emph{structure} of the amplified field. 
If most of the energy is concentrated in tangled multipoles, for instance, it could be that even if fields of order $\sim 10^{14}$~G are produced then still only a small fraction would be available to power precursors. 
It is likely in fact that only a small fraction ($\sim 5\%$; \cite{Reboul-Salze2022MRI}) of the magnetic energy will be deposited into a large-scale dipole.

With respect to the precursor energetics problem, we note that the proposed mechanism for powering the flare requires the crust to fail \cite{tsan12,tsan13}.
As the lattice yields, the failed regions should effectively become plastic rather than elastic \cite{belo14} with the shear modulus turning off there.
This implies that the MRI may now be able to activate in that local area regardless of the magnitude of $S$. 
If the MRI growth time ($\propto \Omega^{-1}$) is faster than the time it takes for energy to escape and the crust to heal, it may still be able to fuel electromagnetic radiation.
While properly assessing this possibility would require simulations of local MRI growth in a small domain encased within a solid and subsequent energy transport, it could be that premerger magnetic amplification still play a r{\^o}le in (possibly rare) cases independently of elastic suppression.

While we have focussed on the MRI in this paper, it should be noted that elasticity will tend to suppress \emph{any} MHD instability that would seek to amplify a seed field due its strictly restorative nature.
This could be relevant to account for in other dynamo scenarios, such as the Tayler-Spruit dynamo \cite{bar22,barr23}, argued to provide a channel for producing magnetars with dipole components exceeding $\gtrsim 10^{14}$~G.
In such cases the star may be spinning up or down rapidly however, and further care is needed as our results assume the crust is in a relaxed state.
As highlighted by \citet{git24} and others, the strain appearing within the tensor \eqref{eq:tauel} will generally not be the Lagrangian one, $\boldsymbol{\xi}$, unless the crust is relaxed prior to the onset of perturbations.
For young (proto-)magnetars, the crust will solidify when the star is still highly oblate and thus the departure may be large.
Even within the context of $\sim$~Gyr old neutron stars taking place in a merger the departure may not be small, as static tides will similarly induce a strain \cite{skk24}.
Additional evolutionary assumptions will be necessary in this case to estimate the extent to which elasticity suppresses instabilities.

How an MRI-produced field may impact any post-merger dynamics -- by amplifying the crustal or even core fields -- is similarly left as an open problem. 
One approach could be to use a simulation setup similar to that from Ref.~\cite{rezz26} but with an inbuilt flag that introduces a strong, turbulent field only $\sim 1$~s prior to merger or so when high-overlap modes become resonantly excited.

\begin{acknowledgments}
AGS thanks Jos{\'e} A. Pons for prompting the question studied in this work following a seminar in Alicante, and the Center of Astrophysics and Gravitation (CENTRA) in Lisbon, the Center of Gravity (CoG) within the Niels Bohr Institute in Copenhagen, and the Albert Einstein Institute (AEI) in Potsdam for hospitality shown while this work was being completed.
{We thank the anonymous referee for their feedback, which improved the quality of this work.}
AGS is grateful for financial support provided by the Deutsche Forschungsgemeinschaft via individual research grant 570901071. 
TC is supported through the Spanish program Unidad de Excelencia María de Maeztu CEX2020-001058-M financed by MCIN/AEI/10.13039/501100011033 and by the MaX-CSIC Excellence Award MaX4-SOMMA-ICE.
AGS and KDK acknowledge funding from the European Union's Horizon MSCA-2022 research and innovation programme ``EinsteinWaves'' under grant agreement No. 101131233.
\end{acknowledgments}


%


\end{document}